\documentclass[journal]{IEEEtran}
\usepackage{amsmath,amsthm,amssymb,paralist,subfigure,cite,graphicx,color}
\usepackage{algorithm2e}
\newtheorem{lem}{Lemma}
\newtheorem{assum}{Assumption}
\newtheorem{theorem}{Theorem}

\newtheorem{cor}{Corollary}

\def\mc{\mathcal}

\begin{document}
\title{Recovering the Structural Observability of Composite Networks via Cartesian  Product}
\author{Mohammadreza Doostmohammadian$^\ast$
\thanks{
$^\ast$ Mechanical Engineering Department, Semnan University, Semnan, Iran \texttt{doost@semnan.ac.ir}.}}
\maketitle

\begin{abstract}
	Observability is a fundamental concept in system inference and estimation. This paper is focused on structural observability analysis of Cartesian product networks. Cartesian product networks emerge in variety of applications including in parallel and distributed systems. We provide a structural approach to extend the structural observability of the constituent networks (referred as the factor networks)
	to that of the Cartesian product network. The structural approach is based on graph theory and is generic. We introduce certain structures  which are tightly related to structural observability of networks, namely parent Strongly-Connected-Component (parent SCC), parent node, and contractions.
	The results show that for particular type of networks (e.g. the networks containing contractions) the structural observability of the factor network can be recovered via Cartesian product. In other words, if one of the factor networks is structurally rank-deficient, using the other factor network containing a spanning cycle family, then the Cartesian product of the two nwtworks is structurally full-rank. We define certain network structures for structural observability recovery. On the other hand, we derive the number of observer nodes--the node whose state is measured by an output-- in the Cartesian product network based on the number of observer nodes in the factor networks. An example illustrates the graph-theoretic analysis in the paper.
	
\end{abstract}

\section{Introduction} \label{sec_intro}
This paper presents a graph-theoretic analysis framework of composite networks formed from smaller constituent networks\footnote{In this paper, graph and network are used interchangeably.} via Cartesian graph product.
Cartesian product of networks has recently emerged in many applications in the literature. One particular application is in the parallel and distributed systems including: multicore systems and multiprocessor System on Chips (SoC) \cite{moraveji2011performance,lin2012embedding},  interconnection
Networks of multicore Chips (NoCs) \cite{flich2008logic}, and parallel routing in computer networks \cite{bakhshi2008efficient}. Cartesian product networks are also prevalent in layered network structures as in piezoelectric  sensor
placement in smart structures \cite{gupta2010optimization}. The other application is in networked dynamic systems \cite{chapman2014controllability,chapmanCDC,carvalho2017composability} where the composite network is the Cartesian product of the sensor network (or the control/actuation network) and the underlying system digraph. Such networked dynamic systems can be found in technological applications such as information networks and power grid monitoring as well as  in biological and social systems.

This paper is focused on structural observability that is a fundamental property of networks. Observability is a measure of how well internal states of the system can be inferred from external outputs (or observations). A system is said to be observable if the entire system states can be determined at any time by tracking the outputs of certain states in finite-time \cite{rugh1996linear}. Therefore, observability is known to be a necessary condition for estimation and filtering \cite{bay,sahu2016distributed,jstsp,wai2016active}. We particularly are interested in  structural observability of network, where each node in the network represents a system state and the links represent the coupling of different states \cite{kronecker_TSIPN}. In this direction, specific graph structures are involved in the structural observability of networks. In another line of research, \cite{nodal_dynamics} claim that the  structural controllability/observability of complex networks are determined by nodal dynamics, not degree distributions. This is particularly important when the system (associated to the network) has pole/zero cancelation. In \cite{zhou2018minimal}, it is proved that as sufficient condition for  observability/controllability of a networked system (or networks of networks), the number of inputs/outputs in each subsystem (or subnetwork) must  be at least equal to that of the maximum \textit{geometric multiplicity} of its state transition matrix. Further, in \cite{zhou2018minimal} it is shown that the observability/controllability of subsystems (or subnetworks) are related to their out/in-degrees.
This paper develops new results on the structural observability of the Cartesian networks based on the structure of its constituent networks. Particularly, the number of observer nodes in the Cartesian network and constituent networks is compared. The literature on this topic is very limited. To the best of our knowledge, only \cite{chapman2014controllability,chapmanCDC,mesbahi2008factorization} present a Gramian-based observability analysis of Cartesian product networks.

\textit{Contributions:} Unlike  \cite{chapman2014controllability,chapmanCDC,mesbahi2008factorization}, here it is shown that the number of sufficient observer nodes for structural observability of composite network  might be less than that of the factor networks. We introduce two types of necessary nodes for structural observability: nodes in parent SCCs and unmatched nodes. It is shown that the structural observability condition for unmatched nodes can be recovered via Cartesian network product, while  for the other the structural observability condition is  stricter and not recoverable. In this direction,  the structural rank of the network is shown to play a key role for structural observability analysis; we discuss certain network structures with structural rank deficiency for which the structural observability is recoverable.  This is significant as in sensor networks and networked dynamic systems the number of observer nodes of certain structures, e.g. star networks, can be reduced by applying proper Cartesian product.

The structural observability analysis in \cite{mesbahi2008factorization} is limited to undirected unweighted networks while in this work directed weighted networks can be considered. The observability approach in this paper is \textit{generic} (or structural) which implies that the results are irrespective of particular link weights and only rely on the network structures. This outperforms the methodologies in the literature, e.g. in \cite{chapman2014controllability,chapmanCDC,mesbahi2008factorization}, as these works provide the observability results for networks representing LTI (Linear-Time-Invariant) systems while the results of this work hold for networks representing LSI (Linear-Structure-Invariant) systems\footnote{For more information on LSI systems refer to \cite{ISJ}.}. In this direction, our observability results are valid for Cartesian product networks with time-variant link weights while the network structures are fixed.

The rest of the paper is organized as follows. In Section~\ref{sec_pre} preliminary notions on graph theory and structural observability as well as problem statement are stated. Section~\ref{sec_aux} presents the introductory results on structural observability of Cartesian network while the main results are discussed in Section~\ref{sec_result}. An illustrative example is given in Section~\ref{sec_ex}. Finally, discussions on the results and concluding remarks are given in Section~\ref{sec_con}.


\section{Preliminary Definitions and Problem Statement} \label{sec_pre}

\subsection{Cartesian Network Product}
\begin{figure}[!h]
	\centering
	{\includegraphics[width=3.2in]{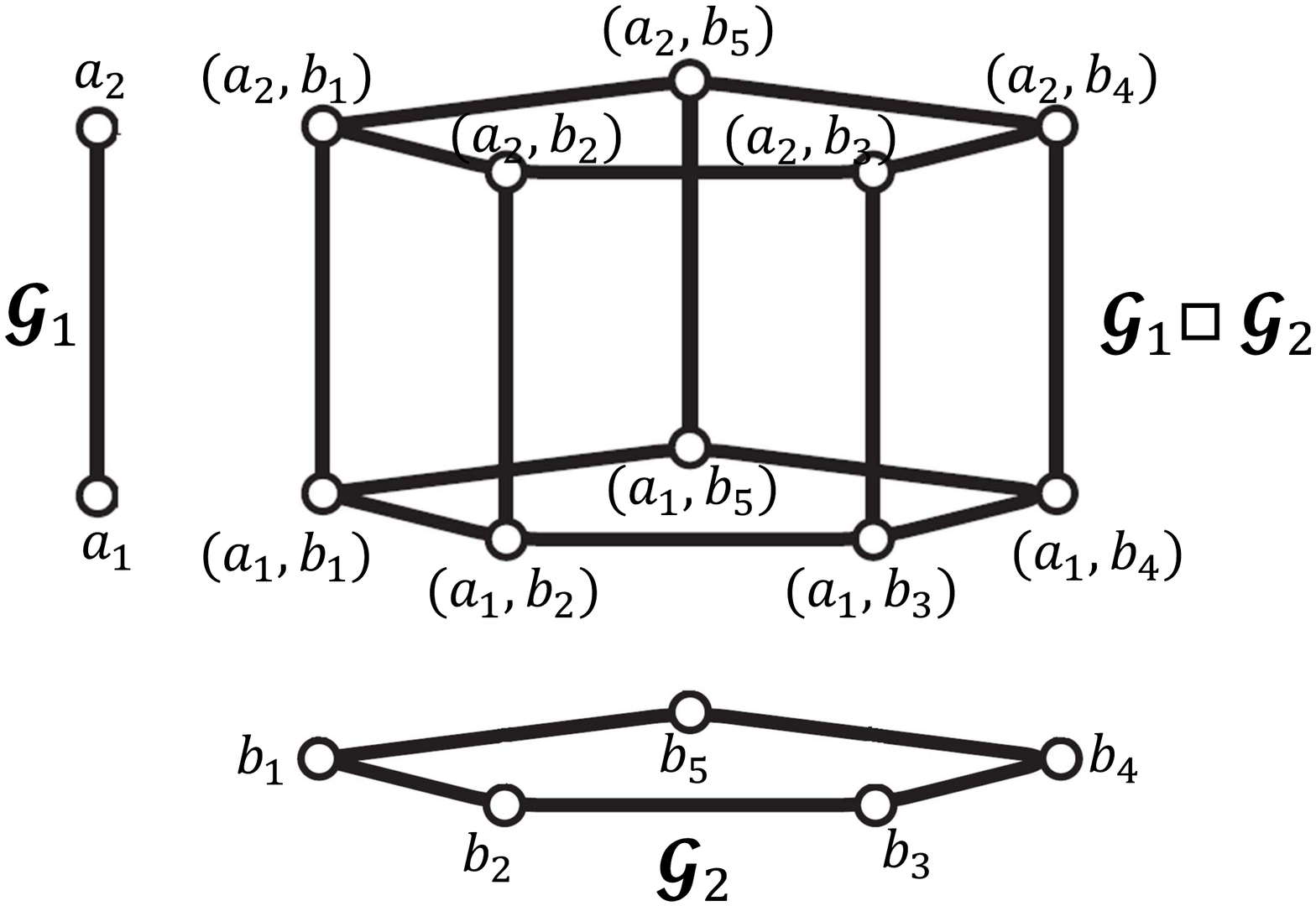}}
	\caption{An example Cartesian product of two  graphs~$\mc{G}_1$ and~$\mc{G}_2$ is shown.}
	\label{fig_cartesian}
\end{figure}

The Cartesian product of two graphs $\mc{G}_1 = (\mc{V}_1,\mc{E}_1)$ and $\mc{G}_2= (\mc{V}_2,\mc{E}_2)$ is a composite graph, denoted as $\mc{G}_C = \mc{G}_1 \Box \mc{G}_2 = \mc{G}_2 \Box \mc{G}_1 $\footnote{Note that unlike other network products, e.g. kronecker product or strong product, the Cartesian product of two graphs is commutative.}, with node set $\mc{V}_C$ and set of links denoted by $\mc{E}_C$. Based on the definition, $\mc{V}_C =\mc{V}_1 \times \mc{V}_2$. In $\mc{E}_C$, two nodes $(a,b)$ and $(a',b')$ are adjacent in the composite graph if $a=a'$, $bb' \in \mc{E}_2$, or $b=b'$, $aa' \in \mc{E}_1$ (see Fig.~\ref{fig_cartesian}). It should be noted that, in this paper, the term $(a,b)$ refers to the node in the Cartesian product network while $ab$ refers to the link from node $a$ to node $b$.  
\begin{equation}
\mc{V}_C = \{(a,b)|a \in \mc{V}_1, b \in \mc{V}_2\}
\end{equation}
\begin{equation}
\mc{E}_C = \{(a,b)(a',b')|a=a', bb' \in \mc{V}_2,~or~ b=b', aa' \in \mc{V}_1\}
\end{equation}
The graphs $\mc{G}_1$ and $\mc{G}_2$ are sometimes referred as the \textit{factor} graphs.

\begin{lem} \label{lem_con}
	The Cartesian product of two graphs is connected if and only if  its  factor graphs are connected. \cite{hammack2011handbook}
\end{lem}
In this paper we make the following assumption:

\begin{assum}
	The factor networks $\mc{G}_1$ and $\mc{G}_2$ are connected. Therefore, from Lemma~\ref{lem_con}, the composite Cartesian network $\mc{G}_C$ is connected.
\end{assum}

\subsection{Structural Observability}
Structural observability recently emerges in the signal processing literature \cite{asilomar11,kruzick2017structurally,commault2011sensor,doostmohammadian2017recovery} as it outperforms the classical Gramian-based observability for the following reasons: (i) it is computationally more efficient, and (ii) the structural analysis is valid for almost-all values of link weights (or structured system parameters) and only depends on the structure of the network.
The main theorem on structural observability of networks is stated below;
\begin{theorem} \label{thm_structural}
	A network is observable if and only if the following two conditions are satisfied:
	\begin{enumerate}[(i)]
		\item There is a directed path from every node to an observer node (and consequently to an output).
		\item There is a disjoint family of cycles and output connected paths spanning all nodes.
	\end{enumerate}
\end{theorem}
\begin{proof}
This theorem is originally proved in \cite{lin, rein_book} for dual problem of structural controllability and here is restated for structural observability.
\end{proof}

\subsection{Problem Statement}
In this paper, given the networks $\mc{G}_1$ and $\mc{G}_2$ the structural observability of the Cartesian network $\mc{G}_C = \mc{G}_1 \Box \mc{G}_2$  is discussed. Particularly, we analyze specific structures and components in the constituent networks  $\mc{G}_1$ and $\mc{G}_2$ that result in unobservable structures in the composite Cartesian network $\mc{G}_C$.
One of the main questions answered in this paper is that if the structural observability of certain structures in $\mc{G}_1$ or $\mc{G}_2$ can be recovered via Cartesian product; and, the follow-up question is: how to recover the structural observability of these structures? These questions are to great extent unexplored in the literature, while the results find significant applications in the structural observability of large-scale Cartesian networks discussed in Section~\ref{sec_intro}.

\section{Related Graph Theroy Notions and Auxiliary Results} \label{sec_aux}
\subsection{Graph Theoretic Observability Notions}
Consider graph $\mc{G}=(\mc{V},\mc{E})$. Define a path from node $a$ to node $b$ as the sequence of directed links in $\mc{E}$ connecting $a$ to $b$, denoted by $a \overset{\scriptsize\mbox{path}}{\rightharpoonup} b$. Denote the mutual path $a \overset{\scriptsize\mbox{path}}{\rightleftharpoons} b$ as $a \overset{\scriptsize\mbox{path}}{\rightharpoonup} b$ and $b \overset{\scriptsize\mbox{path}}{\rightharpoonup} a$. Define a Strongly Connected Component (SCC), denoted by $\mc{S}_i$, as the component in the network where there is a path between every two nodes, i.e. $\forall a,b \in \mc{S}_i: a \overset{\scriptsize\mbox{path}}{\rightleftharpoons} b $. The network $\mc{G}$ is Strongly-Connected (SC) if there is a mutual path between every two nodes in the network, i.e. $\forall a,b \in \mc{V}: a \overset{\scriptsize\mbox{path}}{\rightleftharpoons} b $. Define a parent SCC $\mc{S}^p_i$ as the SCC with no outgoing links to the nodes in other SCCs. In other words, $\nexists ab \in \mc{E}: a \in \mc{S}^p_i, b \notin \mc{S}^p_i$. Denote by $\mc{S}^p$ the set of all parent SCCs. Similarly, define a parent node $\mc{P}_i$ as a node with no outgoing link\footnote{In this work a node with no outgoing links to other nodes in the network while it has a self-cycle is considered as a parent SCC.}, and denote by $\mc{P}$ the set of all parent nodes in the graph.  It should be noted that a parent node/SCC is also referred as \textit{root} node/SCC in the literature \cite{algorithm}.

\begin{lem} \label{lem_one}
	Every connected graph contains at least one parent node or one parent SCC.\footnote{Note that a SC graph is assumed to be one parent SCC.} \cite{algorithm}
\end{lem}

Define the matching, $\underline{\mc{M}}$, as an independent set of links, i.e. no two links in $\underline{\mc{M}}$ share a common end node or common start node. Define a maximum matching, $\mc{M}$, as the matching with maximum possible size. A node is called matched if it is start node of (incident to) a directed link in $\mc{M}$, otherwise the node is called unmatched. The set of unmatched nodes is denoted by $\delta \mc{M}$. Define a cycle family as the set of cycles which are mutually disjoint, i.e. the cycles share no node. Denote by $\mc{C}^m$ the cycle family covering $m$ nodes in the network. If the cycle family covers all $n$ nodes in the network, $\mc{C}^n$, it is called a spanning cycle family.

\begin{lem} \label{lem_cycle}
	A network containing a spanning cycle family has no unmatched node, i.e. for graph $\mc{G}=(\mc{V},\mc{E})$, $|\mc{V}|=n$, if $\exists~\mc{C}^n\subset\mc{E}$ then $\delta \mc{M} = \emptyset$, where $|.|$ represents the cardinality of the set.
\end{lem}
\begin{proof}
	This lemma follows from the fact that every two links belonging to a disjoint cycle family share no common end node or a common start node, and therefore make a  matching. Having the cycle family spanning all nodes, the size of maximum matching is $n$ and, therefore, $\delta \mc{M} = \emptyset$. The detailed illustration can be found in \cite{murota}.
\end{proof}

Next, following the introductory definitions and results in our previous work \cite{asilomar11}, a key theorem on the structural observability of networks is given in the following.

\begin{theorem} \label{thm_obsrv}
	Necessary and sufficient conditions for structural observability of the network are that:
	\begin{enumerate}[(i)]
		\item every unmatched node is observed.
		\item at least one node in every parent SCC is observed.
	\end{enumerate}
\end{theorem}
\begin{proof}
Condition (i) in the above theorem on the structural observability of unmatched nodes for SC networks is proved in \cite{doostmohammadian2018observational,minimal_ISJ}.  Assuming that condition (i) is satisfied, we prove condition (ii) for non-SC networks in the following.

\textit{Necessity:} we prove the necessity by contradiction. Assume no node in parent SCC $\mc{S}^p_i$ is observed. For  structural observability of network to satisfy condition (i) in Theorem~\ref{thm_structural}, for every node there must be a directed path to an observer node.   From the definition of parent SCC, there is no path from the nodes in $\mc{S}^p_i$ to nodes not in $\mc{S}^p_i$ and there is no observer node in $\mc{S}^p_i$. This implies that the condition (i) in Theorem~\ref{thm_structural} for output-connectivity does not hold and the nodes in $\mc{S}^p_i$ are unobservable.

\textit{Sufficiency:} Assume that (at least) one node in every parent SCC is observed. From definition of SCC, there is a path from every node in parent SCC to this observer node. Further, for a non-parent SCC $\mc{S}_j$ there is a path (at least) to one parent SCC $\mc{S}_i^p$; otherwise, $\mc{S}_j$ would be a parent SCC itself. Therefore, there is a directed path from the nodes in $\mc{S}_j$ to the observer node in $\mc{S}^p_i$. This holds for every non-parent SCC. Assuming that the unmatched nodes are observed, having an observation from every parent SCC and following Lemma~\ref{lem_cycle} satisfies both conditions in Theorem~\ref{thm_structural} and the structural observability follows.
\end{proof}

Theorem~\ref{thm_obsrv} finds applications in variety of observability-related research areas including distributed estimation/inference of systems and structural design of observer networks \cite{sahu2016distributed,acc13,kruzick2017structurally,doostmohammadian2017recovery,doostmohammadian2018structural,nuno-suff.ness,complexity_TNSE}.

\subsection{Auxiliary Results on Structural Observability of Cartesian Network}
\begin{figure}[pt]
	\centering
	{\includegraphics[width=3.2in]{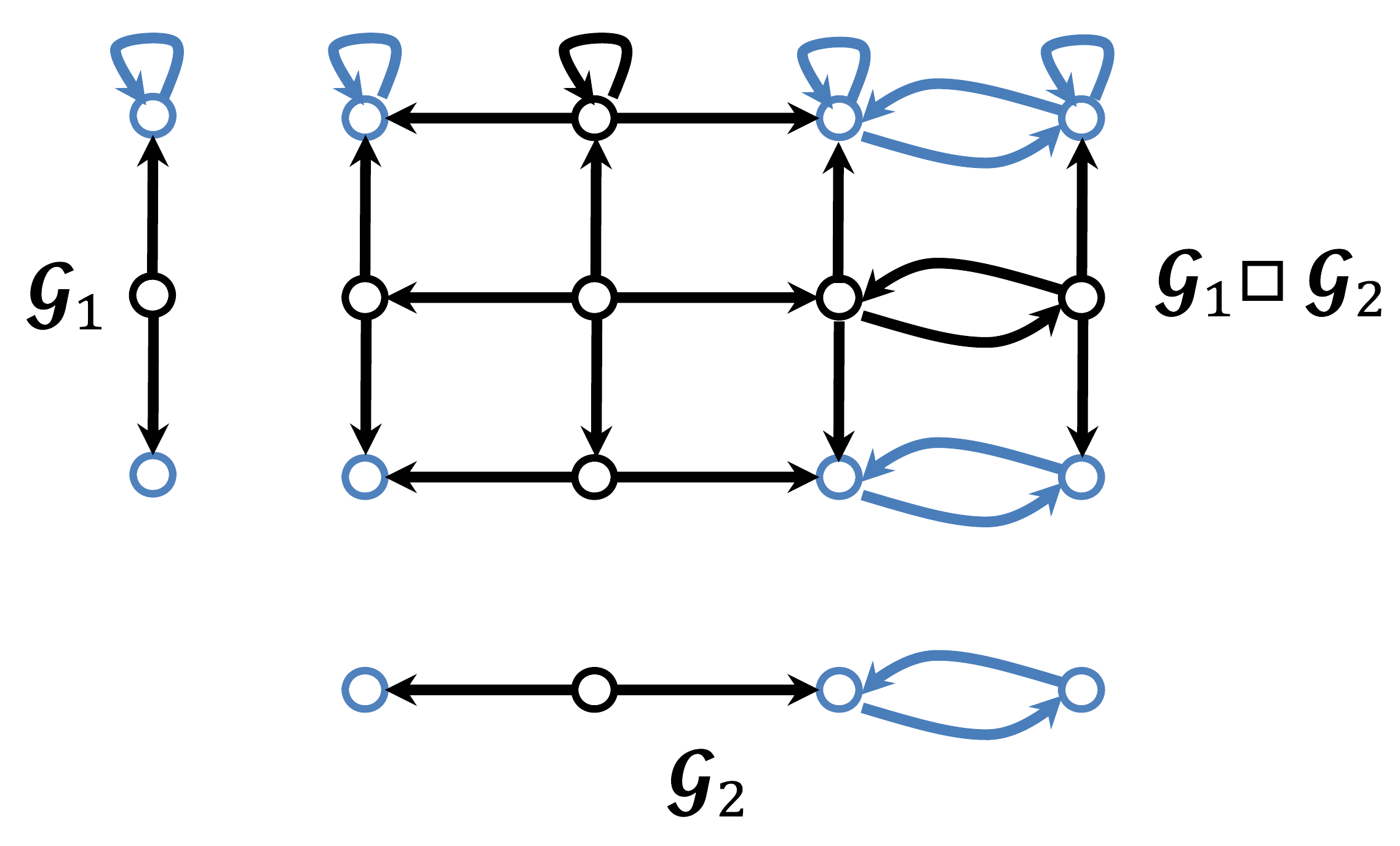}}
	\caption{This figure shows the composite network of two factor networks. Network $\mc{G}_1$ has $1$ parent node and $1$ parent SCC represented in blue.  Network $\mc{G}_2$ also contains a parent node and a parent SCC represented in blue. The parent nodes and parent SCCs in the composite Cartesian network $\mc{G}_1 \Box \mc{G}_2$ are shown in blue. }
	\label{fig_parent}
\end{figure}

In the following, we develop conditions on the structural observability of composite Cartesian network. Fig.~\ref{fig_parent} better illustrates the following lemmas.

\begin{lem} \label{lem_nodeSCC}
If $a$ is a node in $\mc{G}_1$  and  $\mc{S}_j$ is a SCC in $\mc{G}_2$ including the nodes $\{u_1,...,u_l\}$, then the nodes  $\{(a,b)| b \in \{u_1,...,u_l\}\}$ make a SCC in the composite product graph  $\mc{G}_C$.\footnote{Note that because of the commutative property of the Cartesian product, the same result can be stated for a node in $\mc{G}_2$ and a SCC in $\mc{G}_1$. This commutative property holds for all the lemmas and theorems in this paper.}
\end{lem}
\begin{proof} The proof of this lemma follows the definition of the Cartesian product and SCC. Based on the definition of SCC for every two  nodes $u_m$ and $u_n$ in $\{u_1,...,u_l\}$, we have $u_m \overset{\scriptsize\mbox{path}}{\rightleftharpoons} u_n$. This  implies that for every two nodes $(a,u_m)$ and $(a,u_n)$ in the composite graph belonging to the set  $\{(a,b)| b \in \{u_1,...,u_l\}\}$, we have $(a,u_m) \overset{\scriptsize\mbox{path}}{\rightleftharpoons} (a,u_n)$, and the lemma follows.
\end{proof}
 	
\begin{lem}\label{lem_SCC}
If $\mc{S}_i$ is a SCC in $\mc{G}_1$ and  $\mc{S}_j$ is a SCC in $\mc{G}_2$, then the nodes  $\{(a,b)|a \in \mc{S}_i, b \in \mc{S}_j\}$ make a SCC in the composite product graph  $\mc{G}_C$.
\end{lem}
\begin{proof} Let $(a,b)$ and $(a',b')$ be two nodes in $\mc{G}_C$ where $a,a' \in \mc{S}_i$ and $b,b' \in \mc{S}_j$. We need to prove that for every two such nodes we have $(a,b) \overset{\scriptsize\mbox{path}}{\rightleftharpoons} (a',b')$. If $a=a'$ or $b=b'$, the proof simply follows from Lemma~\ref{lem_nodeSCC}; otherwise, from Lemma~\ref{lem_nodeSCC} we have $(a,b) \overset{\scriptsize\mbox{path}}{\rightleftharpoons} (a',b)$ since $a,a' \in \mc{S}_i$ and $(a',b) \overset{\scriptsize\mbox{path}}{\rightleftharpoons} (a',b')$ since $b,b' \in \mc{S}_j$. Therefore, from definition of a path we have $(a,b) \overset{\scriptsize\mbox{path}}{\rightleftharpoons} (a',b')$ and the lemma follows.
\end{proof}
	
\begin{lem}\label{lem_parentnode}
	For every parent SCC $\mc{S}^p_i$ and parent node $\mc{P}_j$ in  $\mc{G}_1$ and a parent node $\mc{P}_k$ in $\mc{G}_2$, respectively, there is a parent SCC and a parent node in the Cartesian product graph  $\mc{G}_C$.
\end{lem}
\begin{proof} Following the definition of parent node for $\mc{P}_j$ and $\mc{P}_k$, there is no outgoing link from nodes $(\mc{P}_j,b)$ to other nodes $\{(a,b)|a \neq \mc{P}_j\}$ and from nodes $(a,\mc{P}_k)$ to other nodes $\{(a,b)|b \neq \mc{P}_k\}$ in $\mc{G}_C$. Therefore, there is no outgoing link from  $ (\mc{P}_j,\mc{P}_k)$ to nodes $\{(a,b)|a \neq \mc{P}_j,b \neq \mc{P}_k\}$ implying that $ (\mc{P}_j,\mc{P}_k)$ is a parent node. Further, from Lemma~\ref{lem_nodeSCC} the nodes $\{(a,\mc{P}_k)|a \in \mc{S}_i^p \}$ make a SCC in $\mc{G}_C$. From the definition of parent node and parent SCC there is no link from $\{(a,\mc{P}_k)|a \in \mc{S}_i^p \}$ to nodes $\{(a,\mc{P}_k|a \notin \mc{S}_i^p \}$ and to nodes $\{(a,b)|b \neq \mc{P}_k\}$ implying that $\{(a,\mc{P}_k)|a \in \mc{S}_i^p \}$ is a parent SCC.
\end{proof}
	
\begin{lem}\label{lem_parent}
	For every parent SCC $\mc{S}^p_i$ and parent node $\mc{P}_j$ in  $\mc{G}_1$ and a parent SCC $\mc{S}^p_k$ in $\mc{G}_2$ there is a parent SCC in the composite Cartesian graph  $\mc{G}_C$.
\end{lem}
\begin{proof}
Following the commutative property of Cartesian product and Lemma~\ref{lem_parentnode}, the nodes $\{(\mc{P}_j,b)|b\in \mc{S}^p_k\}$ represent a parent SCC in $\mc{G}_C$. Further, from Lemma~\ref{lem_SCC} the nodes $\{(a,b)|a \in \mc{S}^p_i, b\in \mc{S}^p_k\}$ make a SCC in  $\mc{G}_C$. For every node $\{(a,b)|a \in \mc{S}^p_i \}$ there is no link to nodes $\{(a,b)|a \notin \mc{S}^p_i \}$. Similarly, for nodes $\{(a,b)|b \in \mc{S}^p_k \}$ there is no link to nodes $\{(a,b)|b \notin \mc{S}^p_k \}$. Therefore, the SCC $\{(a,b)|a \in \mc{S}^p_i, b\in \mc{S}^p_k\}$ has no outgoing link to nodes $\{(a,b)|a \notin \mc{S}^p_i, b\notin \mc{S}^p_k\}$ and the lemma follows.
\end{proof}
	
An immediate corollary of the above lemmas and proofs is stated in the following:

\begin{cor} \label{cor_parent}
	For every parent node and parent SCC in the factor graphs $\mc{G}_1$ and $\mc{G}_2$ there is one and only one parent node or parent SCC in the composite graph $\mc{G}_C$.
\end{cor}

\begin{lem}\label{lem_parent_matched}
	Every parent node $\mc{P}_i$ is unmatched, i.e. $\mc{P}_i \in \delta \mc{M}$.
\end{lem}
\begin{proof} From the definition, every matched node in $\mc{V}\backslash \delta \mc{M}$ is the start node of a directed link in a maximum matching $\mc{M}$. From the definition of parent node $\mc{P}_i$, it has no outgoing link  and therefore it is not the start node of any link. This implies that every parent node is unmatched, i.e. $\mc{P}_i \in \delta \mc{M}$.
\end{proof}

\section{Main results} \label{sec_result}
Using the introductory lemmas in the previous sections, this section provides the main results of the paper.

\begin{figure}[!h]
	\centering
	{\includegraphics[width=2.6in]{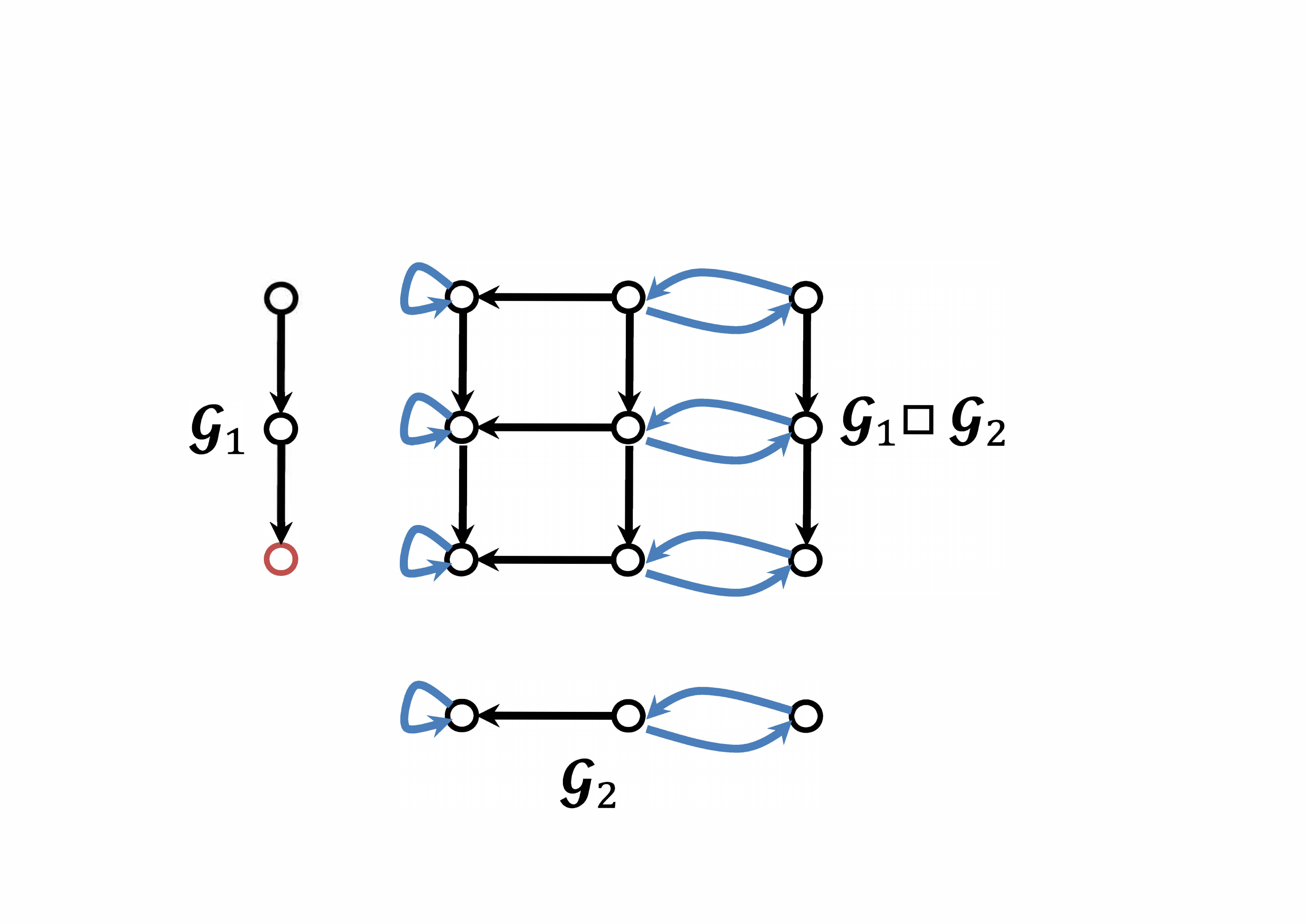}}
	\caption{The composite network of two factor networks $\mc{G}_1$ and $\mc{G}_2$ is shown. Network $\mc{G}_1$ has no spanning cycle family and contains an unmatched node represented in red. Network $\mc{G}_2$ contains a spanning cycle family $\mc{C}^3$ represented in blue. Following Theorem~\ref{thm_unmatched}, the composite network includes a spanning cycle family $\mc{C}^9$ and therefore has no unmatched nodes. }
	\label{fig_cycle}
\end{figure}

\begin{theorem} \label{thm_unmatched}
    Let one of the factor graphs $\mc{G}_1$ or $\mc{G}_2$ contains a spanning cycle family. Then in the composite Cartesian graph $\mc{G}_C$ we have $\delta \mc{M}_C = \emptyset$.
\end{theorem}
\begin{proof}
Assume that the graph $\mc{G}_1$ contains a spanning cycle family $\mc{C}^n$ and for graph $\mc{G}_2$ we have $\delta \mc{M}_2 \neq \emptyset$. Consider node $(a,b)$ in the composite graph $\mc{G}_C$; for every $b$ associated with $\mc{G}_2$ we have $a \in \mc{C}^n$ and therefore in $\mc{G}_C$ we have $(a,b) \in \mc{C}^n$ (see Fig.~\ref{fig_cycle}). In other words, for every node $b$ in $\mc{G}_2$ all the nodes associated with the graph $\mc{G}_1$ are included in a cycle family $\mc{C}^n$ in the graph $\mc{G}_C$, i.e. $\forall b \in \mc{G}_2, (a,b) \in \mc{C}^n$. Assume the size of $\mc{G}_2$ to be $N$. Therefore, we have $N$ set of disjoint cycle family $\mc{C}^n$ in $\mc{G}_C$ which implies that $\exists~\mc{C}^{Nn} \subset \mc{E}_C$. From Lemma~\ref{lem_cycle}, having a spanning cycle family  $\mc{C}^{Nn}$ over all the $Nn$ nodes in $\mc{G}_C$ implies that $\delta \mc{M}_C = \emptyset$.
\end{proof}
	
\begin{cor} \label{cor_unmatched}
	Having $\delta \mc{M}_1 \neq \emptyset$ in network $\mc{G}_1$, then
	Theorem~\ref{thm_unmatched} implies that a network $\mc{G}_2$ with a spanning cycle family recovers the condition (ii) in Theorem~\ref{thm_structural} for structural observability of the Cartesian product network $\mc{G}_C$.
\end{cor}
This corollary implies that one can recover the structural observability of certain factor networks having unmatched nodes by applying Cartesian product of another  network containing a  spanning cycle family.

\begin{theorem} \label{thm_parent}
	Let $|\mc{S}^p_1 \cup \mc{P}_1|$, $|\mc{S}^p_2 \cup \mc{P}_2|$, and $|\mc{S}^p_C \cup \mc{P}_C|$ represent the total number of parent SCCs and parent nodes respectively in the graphs $\mc{G}_1$, $\mc{G}_2$ and the composite Cartesian graph $\mc{G}_C$. Then,  $|\mc{S}^p_C \cup \mc{P}_C| = |\mc{S}^p_1 \cup \mc{P}_1|\times|\mc{S}^p_2 \cup \mc{P}_2|$.
\end{theorem}
\begin{proof}
From Lemma~\ref{lem_parentnode}, Lemma~\ref{lem_parent}, and Corollary~\ref{cor_parent}, for every parent node and parent SCC in the factor networks $\mc{G}_1$ and $\mc{G}_2$ there is only one parent node/SCC in the network $\mc{G}_C$. This implies that for each parent node/SCC in graph $\mc{G}_1$ there are $|\mc{S}^p_2 \cup \mc{P}_2|$ parent node/SCC associated with the graph $\mc{G}_2$ in composite graph $\mc{G}_C$ (see Fig.~\ref{fig_parent}). Therefore, the total number of parent nodes and parent SCCs in $\mc{G}_C$ is equal to $|\mc{S}^p_1 \cup \mc{P}_1|\times|\mc{S}^p_2 \cup \mc{P}_2|$ and the theorem follows.
\end{proof}
Following the above theorem, Lemma~\ref{lem_one}, and Lemma~\ref{lem_parent}, the following corollary holds.
\begin{cor} \label{cor_parentSCC}
	If  any of the factor networks has (at least) one parent SCC, then the Cartesian product network $\mc{G}_C$ has (at least) one parent SCC, and therefore, an observer node from the parent SCC is required for structural observability of $\mc{G}_C$. 
\end{cor}
It should be noted the parent nodes are indeed unmatched nodes (see Lemma~\ref{lem_parent_matched}). Therefore, from Theorem~\ref{thm_unmatched}, their structural observability in the composite network $\mc{G}_C$ can be recovered via Cartesian product of another network with spanning cycle family. However,  Corollary~\ref{cor_parentSCC} implies that the structural observability of a parent SCC in a factor network cannot be recovered in the Cartesian product network $\mc{G}_C$.  The results of this section are not stated in the literature \cite{chapman2014controllability,chapmanCDC,mesbahi2008factorization}.

\begin{figure}[!h]
	\centering
	{\includegraphics[width=3in]{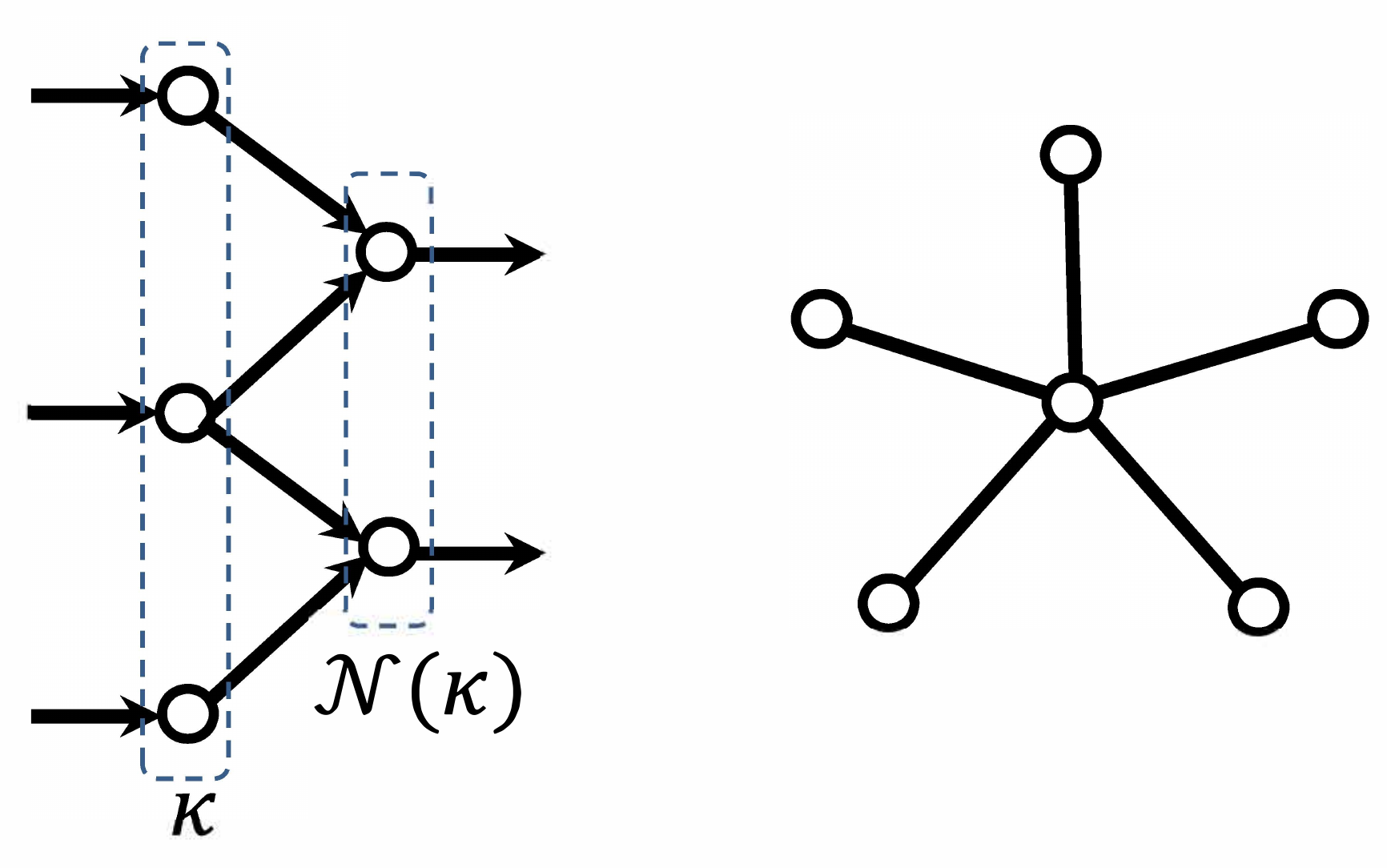}}
	\caption{(Left) the contraction in a directed network is shown. The subset of nodes $\kappa$ are contracted/linked to their neighboring subset of nodes $\mc{N}(\kappa)$ where $|\mc{N}(\kappa)|<|\kappa|$. Such a contraction contains an unmatched node in $\kappa$. (Right) A star network is shown which contains contractions and unmatched nodes. }
	\label{fig_cont_star}
\end{figure}

Based on Theorem~\ref{thm_unmatched} and Corollary~\ref{cor_unmatched}, the structural observability of the networks with unmatched nodes can be recovered. In this direction, we introduce particular network structures that contain unmatched observer nodes.
\begin{lem}
	The number of unmatched observer nodes in the network can be determined based on the structural-rank\footnote{The structural rank is defined as the maximum possible rank of the network adjacency matrix $A$ by changing the link weights (non-zero entries of $A$). In other words, the structural rank of the adjacency matrix is the number of distinct non-zero entries that share no rows and columns.} (or S-rank) of the network adjacency matrix. For a network of size $n$ and adjacency matrix $A$,
	\begin{equation}
	|\delta \mc{M}| = n - \mbox{S-rank}(A)
	\end{equation}
\end{lem}
This lemma is discussed in the previous work by the author \cite{doostmohammadian2018observational}. Some  network structures result in larger S-rank deficiency and more unmatched nodes. An example of such network structures is a \textit{contraction}. Roughly speaking, a contraction is defined as a graph component in which more nodes are linked (contracted) to less other nodes. For graph $\mc{G}=(\mc{V},\mc{E})$, define a contraction  as the subset of nodes $\kappa$ such that $|\mc{N}(\kappa)|<|\kappa|$ where $\mc{N}(\kappa)=\{b|ab \in \mc{E}, a \in \kappa\}$. For more details on these components refer to \cite{doostmohammadian2018observational}. Such components can be found, for example, in \textit{star graphs}. See Fig.~\ref{fig_cont_star} for examples of contraction and star networks. More examples of S-rank deficient networks are provided in \cite{icassp13}.

\section{Illustrative Example} \label{sec_ex}
\begin{figure}[!h]
	\centering
	{\includegraphics[width=3.4in]{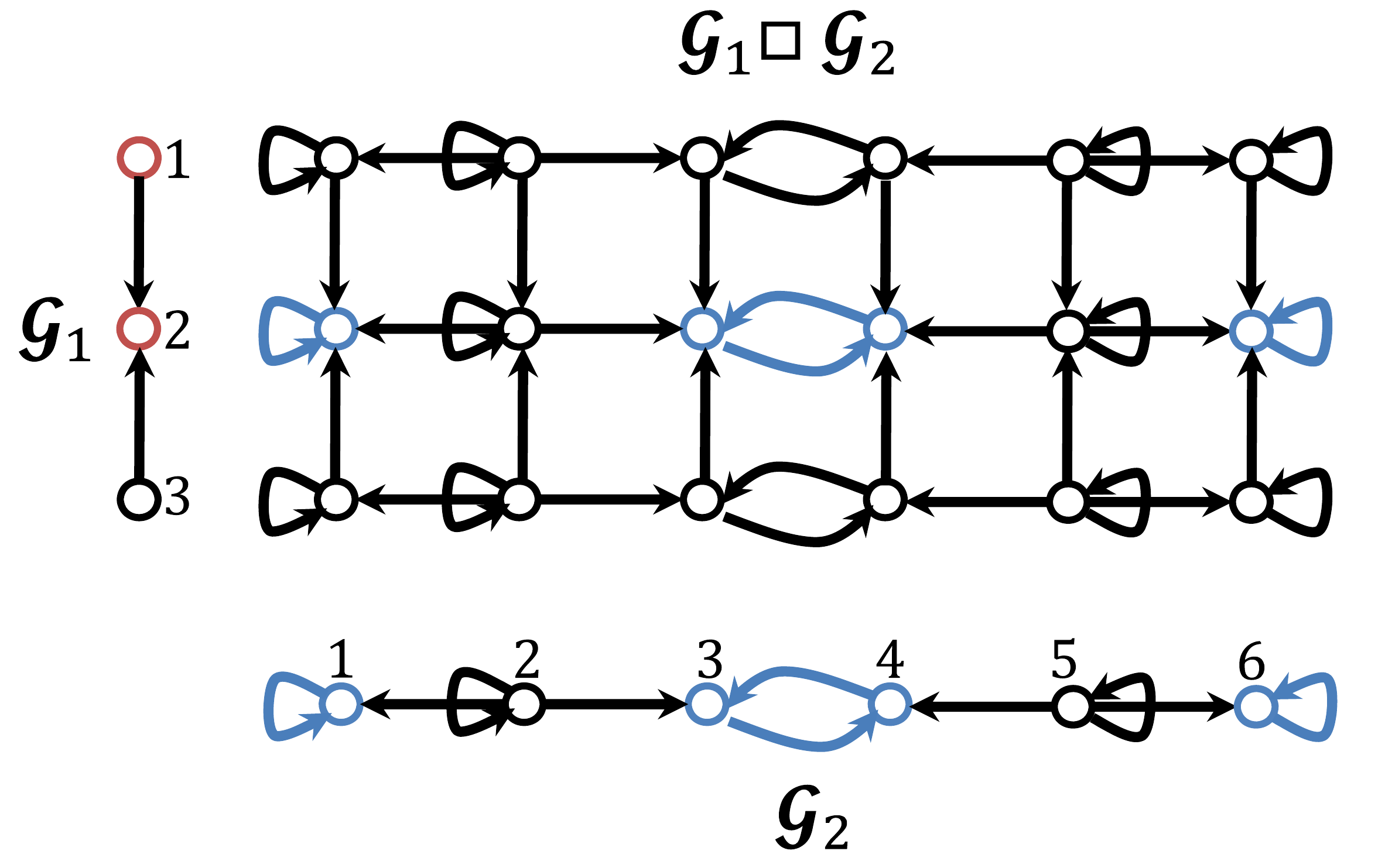}}
	\caption{This figure shows the Cartesian product of two graphs $\mc{G}_1$ and $\mc{G}_2$. Network $\mc{G}_1$ represents a contraction and includes $2$ unmatched nodes represented in red (one parent node). Network $\mc{G}_2$ includes $3$ parent SCCs represented in blue and a spanning cycle family $\mc{C}^6$. The resulting Cartesian product network has $3$ parent SCC represented in blue. }
	\label{fig_example}
\end{figure}

This section provides an example to further illustrate the results of the previous sections. Consider Cartesian product of two networks shown in Fig.~\ref{fig_example}.
This graph may represent the structure of a NoCs as described in \cite{flich2008logic}. Network $\mc{G}_1$ represents a contraction of $2$ nodes into $1$ node. One possible set of links for maximum matching in $\mc{G}_1$ is $\mc{M}=\{32\}$ and the set of unmatched nodes is $\delta \mc{M} = \{1,2\}$. Note that the maximum matching and therefore the set of unmatched nodes is  not unique in general. For example, in $\mc{G}_1$ either of nodes $1$ or $3$ could be unmatched and these two nodes are observationally equivalent (see \cite{doostmohammadian2018observational}  for more information).  The network $\mc{G}_2$ includes a spanning cycle family $\mc{C}^6=\{\{11\},\{22\},\{34,43\},\{55\},\{66\}\}$. The set of parent SCCs in $\mc{G}_2$ is $\mc{S}^p = \{\{1\},\{3,4\},\{6\}\}$. According to Theorem~\ref{thm_unmatched}, the Cartesian product network $\mc{G}_C=\mc{G}_1 \Box \mc{G}_2$ contains a spanning cycle family $\mc{C}^{18}$ and has no unmatched nodes. The number of parent SCCs in $\mc{G}_C$ follows Theorem~\ref{thm_parent}. There is $1$ parent node in $\mc{G}_1$ and $3$ parent SCCs in $\mc{G}_2$; therefore, $|\mc{S}^p_C \cup \mc{P}_C|=3$. Since $\mc{G}_C$ contains a spanning cycle family, there is no unmatched node and no parent node in $\mc{G}_C$. Therefore, $|\mc{S}^p_C|=3$ as shown in Fig.~\ref{fig_example}. For structural observability, following Theorem~\ref{thm_obsrv}, one node in every parent SCC in $\mc{G}_C$ must be observed. Making observations from nodes $(2,1)$, $(2,3)$, and $(2,6)$ makes $\mc{G}_C$ observable. Note that the spanning cycle family in $\mc{G}_2$ recovers the condition (i) in Theorem~\ref{thm_obsrv} for structural observability of unmatched nodes in $\mc{G}_2$. Therefore, only the observation from parent SCCs result in structural observability of the Cartesian product network $\mc{G}_C$. To check the structural observability we run Kalman estimation \cite{kalman:61} over the network.  Assume the network to simulate a  system $\dot{x}=Ax+v$, where $x$ represents the state of each node, $A$ is the adjacency matrix of the composite network $\mc{G}_C$ defining the system dynamics, and $v=\mc{N}(0,0.05)$ represents the Gaussian noise term.  The entries of $A$ (the link weights in $\mc{G}_C$) are chosen randomly. Having the given observations, the rank of the Gramian matrix also checked as follows:
\begin{eqnarray}
   \mbox{rank}\left( \begin{array}{c}
   C \\
   CA \\
   \vdots \\
   CA^{17}
   \end{array}\right) = 18.
\end{eqnarray}

\begin{figure}[t]
	\centering
	{\includegraphics[width=3.4in]{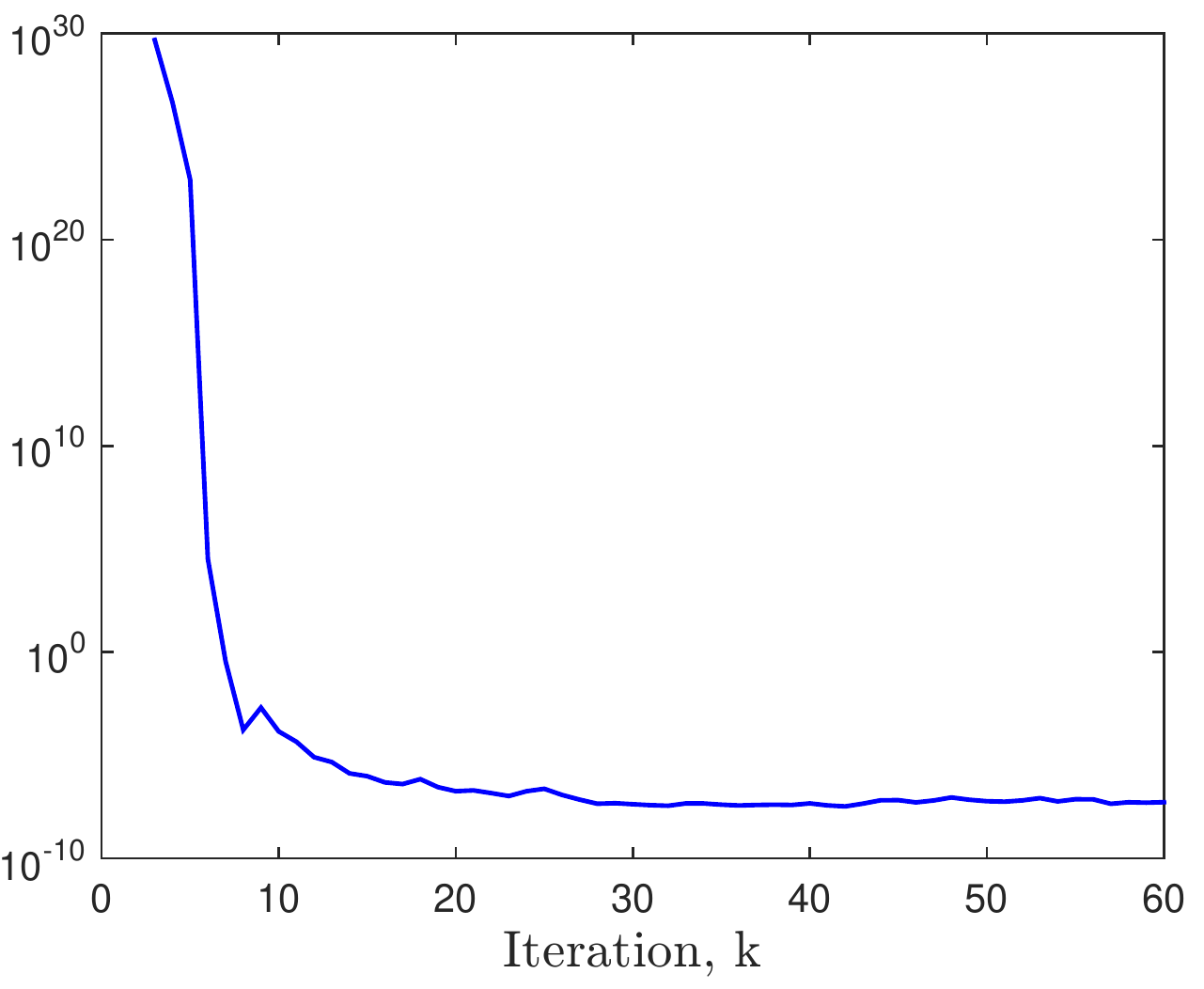}}
	\caption{This figure shows the time evolution of MSEE over the system dynamics represented by the network of Fig.~\ref{fig_example}. The estimation protocol is a simple open-loop least-square estimator based on inverting the observability Gramian matrix. The MSEE is bounded steady-state stable implying the structural observability of the underlying system and therefore the Cartesian network in Fig.~\ref{fig_example} is observable.  }
	\label{fig_sim}
\end{figure}

The condition number, i.e. the square root of the ratio of the largest eigenvalue of observability Gramian to its smallest, is $48.55$.
To avoid trivial solution for tracking the system dynamics, the entries of $A$ are chosen such that $\rho (A)>1$, where $\rho(A)$ represents the spectral radius of  matrix $A$. This implies that the network represents a potentially unstable system dynamics. The initial state values $x(0)$ are chosen randomly in the range $[-5,5]$, and the noise in the  observations is $r=\mc{N}(0,0.05)$. The simulation result for open-loop least-square estimator (based on inverting the Gramian matrix) over $\mc{G}_C$ is shown in Fig.\ref{fig_sim}.
The Mean-Squared Estimation Error (MSEE) for tracking $18$ states are shown over time averaged over $100$ independent Monte-Carlo trials. The bounded steady-state stable MSEE is a result of the  structural observability of network while the network dynamics is chosen to be unstable ($\rho(A)>1$).

It should be noted that the observability of the network in Fig.~\ref{fig_example} is irrespective of link weights and only rely on the network structure. This is in contrast to works \cite{chapman2014controllability,chapmanCDC,mesbahi2008factorization} where the observability analysis relies on numerically checking  the rank of the Gramian matrix.
In general, for structural observability, there are graph-theoretic algorithms to find the parent SCCs and unmatched nodes in large-scale networks. \textit{Depth-First-Search (DFS)} algorithm finds the SCCs and their partial order in a network \cite{algorithm}. The algorithm is also known as \textit{Tarjan} algorithm \cite{tarjan}. This algorithm is of polynomial-order complexity $\mc{O}(n^2)$. Further, the \textit{Hopcroft-Karp} algorithm (also known as \textit{Dulmage-Mendelsohn decomposition}) finds one possible maximum matching and set of unmatched nodes in the network by computational complexity $\mc{O}(n^{2.5})$ \cite{hopcraft}.

\section{Discussions and Concluding Remarks} \label{sec_con}
This paper investigates the structural observability properties of the Cartesian network products based on the structural characteristics of its factor (constituent) networks. Two main conditions for  structural observability of network are analyzed over the Cartesian network: observing (i) the unmatched nodes, and (ii) the parent SCCs. It is shown that if one of the factor networks contains a spanning cycle family the structural observability of unmatched nodes in other network is recovered. Further, the number of parent SCCs/nodes in the Cartesian product network is derived based on the number of parent SCCs/nodes in the factor networks. The results show that if any of the factor networks contains a parent SCC the Cartesian network also contains a parent SCC, and therefore the structural observability of parent SCCs is not recoverable. However, for S-rank deficient networks with unmatched nodes the structural observability can be recovered. Therefore, in case one of the factor networks can be designed by the user (for example in design of parallel and distributed systems \cite{moraveji2011performance,lin2012embedding,flich2008logic,bakhshi2008efficient}), the factor network can be designed such that the overall Cartesian network requires less observer nodes.

The structural observability properties of $\mc{G}_C = \mc{G}_1 \Box \mc{G}_2$ can be deduced from structural observability of the factor networks  $\mc{G}_1$ and $\mc{G}_2$. Particularly, Theorem~\ref{thm_unmatched} implies that if $\mc{G}_1$ or $\mc{G}_2$ includes a spanning cycle family, then the composite Cartesian network $\mc{G}_C$ is structurally full rank and has no unmatched node implying that the condition (i) in Theorem~\ref{thm_obsrv} is satisfied. Further, the number of parent SCCs in $\mc{G}_C$ can be determined by the number of parent SCCs and parent nodes in $\mc{G}_1$ and $\mc{G}_2$ based on Theorem~\ref{thm_parent}. The number of parent SCCs and unmatched nodes determine the structural observability  of the composite network. Therefore, the structural results of this paper imply that one can assess the structural observability of Cartesian product network without explicitly constructing it, and thus, one may save significant computational effort.

The structural results in this paper are based on  graph-theoretic algorithms as discussed in Section~\ref{sec_ex}. These algorithms are computationally more efficient over numerical approach in the literature as in \cite{chapman2014controllability,chapmanCDC,mesbahi2008factorization}. These papers are based on the observability Gramian with the computational complexity of (at least) $\mc{O}(n^3)$,  while the structural approach of this paper is of overall computational complexity $\mc{O}(n^{2.5})$. The better computational efficiency is particularly preferable in large-scale applications.

The observability results in this paper are \textit{generic}, in the sense that they hold for almost all numerical values of network link weights\footnote{It is known that the numerical values for which the generic property does not hold lies on algebraic subspace with \textit{zero Lebesgue measure} \cite{woude:03}.}, even if the link weights are time-variant. The networks may represent a linear system or linearization of a nonlinear system \cite{liu_pnas,haber2018state}. For example, for linear system, $\dot{x} = Ax$, matrix $A$ is the network adjacency matrix. The matrix $A$ could be time-variant in the sense that its zero-nonzero pattern (representing the network structure) is fixed while the link weights change in time. This is also the case in linearization of nonlinear systems, where the structure of the system Jacobian $\mc{J}$ is fixed while the entries are function of the linearization point. Assume, the network models a nonlinear system as $\dot{x} = f(x)$; in the linearized model, the Jacobian entries $\mc{J}_{ij}=\frac{\partial f_i}{\partial x_j}$ are nonzero if $f_i$ is a function of $x_j$, otherwise $\mc{J}_{ij}$ is fixed zero. The numerical value of nonzero entries of $\mc{J}$ depends on the operating (linearization) point, however the zero-nonzero pattern of $\mc{J}$ is irrespective of the operating point \cite{nonlin}. Therefore, the structural
observability results are valid for linearization of nonlinear systems \cite{haber2018state}.

\bibliographystyle{IEEEbib}
\bibliography{bibliography}


\end{document}